# Discrete chiral ballistic polariton laser

Zuzanna Werner,[1, *] Andrzej Frączak,[1] Valtýr Kári Daníelsson,[2]
Jacek Szczytko,[1] Barbara Piętka,[1] and Helgi Sigurðsson[1, 2, †]

[1] *Institute of Experimental Physics, Faculty of Physics, University of Warsaw, ul. Pasteura 5, PL-02-093 Warsaw, Poland*
[2] *Science Institute, University of Iceland, Dunhagi 3, IS-107 Reykjavik, Iceland*

Orbital angular momentum (OAM) of light appears when the phase of an electromagnetic wavefront winds around its direction of propagation, also known as optical vorticity. Contrary to the binary-valued photon spin, the integer-valued optical vortex charge is unbounded with many advantages in optical communication and trapping and enhancing the capacity of data encoding and multiplexing. Singular optoelectronic and chiroptic quantum technologies rely on the development of coherent and compact light sources of well-defined and reconfigurable OAM. We propose an optically tunable *discrete chiral exciton-polariton microlaser* that leverages strong spin-dependent polariton interactions, structured pumping, and inherent cavity photon spin-to-angular momentum conversion to emit coherent nonlinear light of variable OAM. By choosing pumping patterns with broken inversion symmetry in the microcavity plane, we invoke geometric frustration between spinor ballistic condensates which spontaneously obtain a high-charge circulating current locked with the pump polarization. Our optically configurable system requires only a planar cavity, thus avoiding the need for specialized irreversible cavity patterning or metasurfaces.

## I. INTRODUCTION

Controlling the spin and orbital degrees of freedom of photons in coherent light sources forms an important step in advancing chiral optical technologies. The flexible generation of unbounded and variable orbital angular momentum (OAM) states of vectorial light, or optical vortices [1], within compact lasing devices finds application in many optical and quantum technologies such as optical trapping, sensing, communication and information processing, optomechanics, and strong light-matter interactions [2–4]. Nanopatterned metasurfaces [5–7], ring resonators [8, 9], or Archimedean spiral gratings [10] are some of the most prominent strategies to achieve chiral operation and vortex lasing [2] but can become impeded by the lack of tunability due to their specific and irreversible fabrication demands, often precluding in-situ control.

Here, we propose an all-optically tunable high-charge discrete vortex microlaser based on spin orbit coupled (SOC) ballistic exciton-polariton condensates. Exciton-polaritons (from here on, *polaritons*) are half-light and half-matter quasiparticles appearing in the strong coupling regime between excitons and confined photons, typically in planar semiconductor microcavities [11]. They possess two integer spin projections $s = \pm 1$ along the cavity optical axis, explicitly related to the circular polarization $\sigma^{\pm}$ of the cavity photon. Being composite bosons, they can undergo a power-driven nonequilibrium phase transition into a macroscopic coherent quantum state at elevated temperatures referred to as a polariton condensate [11] described by a spinor order parameter $\Psi \equiv (\psi_+, \psi_-)^{\mathrm{T}}$. In their condensed form, strong interactions inherited from the excitons imbue polaritons with superfluid properties [12, 13] and the possibility to form quantized vortices [14, 15]. The macroscopic coherence of the condensate is directly transferred to the cavity emission, underpinning the concept of low threshold inversionless polariton lasers [16], where quantized polariton vortices become sources of coherent optical vortex beams.

Polariton vorticity can manifest within excited states of transverse confining potentials in planar microcavities, such as cavity mesas [17, 18], open cavities [19], single [20] and multiple micropillars [21, 22], accidental defects [23], and optical traps [24–30]. In this context, polariton systems have garnered interest as low threshold chiral lasing devices [20, 31, 32] because of their large optical nonlinearities, susceptibility to photonic SOC inherent to microcavities [21, 33] in particular those imbued with nematic liquid crystals [34, 35], and with the rising amount of materials for room temperature condensation [36].

Under nonresonant excitation, trapped polariton condensates typically display spontaneous formation of single-charge [18–20, 23, 25, 26, 29] or half-charge [37–39] polariton vortices while theoretical studies have shown that high charge vortices can also spontaneously form above threshold [40, 41]. High charge vortices are instead achieved using resonant excitation schemes wherein the phase profile of an optical vortex is imprinted onto the polariton condensate [42–46]. In particular, resonant excitation techniques were used to explore the interplay between Rabi vortices and the polariton spinor order parameter revealing rich spin physics including full Poincaré beams [46]. However, for a nonresonantly pumped polariton system, the spontaneous formation of condensates carrying high-charge vortices locked with the spin angular momentum (SAM) of the pump, arising from the balance of gain and losses, has been scarcely explored.

Recently, giant discrete vortices with all-optically tun-able high-charge OAM were reported in systems of interacting ballistic polariton condensates [47]. By tailoring the excitation laser profile into odd-numbered discrete rotational (polygonal) geometries, breaking in-plane inversion symmetry, the condensates could be forced into states carrying persistent high-charge circulating currents. These states are similar to optical vortex solitons [48] appearing in coupled nonlinear waveguides [49], phase-locked laser arrays [50], or photonic lattices [51]. Here, we show that the combination of structured spin-polarized pumping and photonic SOC locks the OAM of these giant vortex condensates with the SAM of the pump, facilitating an optically tunable high-charge vortex microlaser [see schematic Fig. 1(a)] and exceptional points. We draw a qualitative comparison between structured and non-structured pumping and demonstrate how an excitation profile of discrete rotational symmetry offers multiple, and previously unexplored, pump parameters to variably change OAM of the emission. The ballistic nature of the condensates leads to rich patterns of polarization singularities such as V-point and C-points [52] and opens perspectives on using discrete polariton vortices for high order Poincaré beam generation [53].

* z.werner@student.uw.edu.pl
† helgi.sigurdsson@fuw.edu.pl

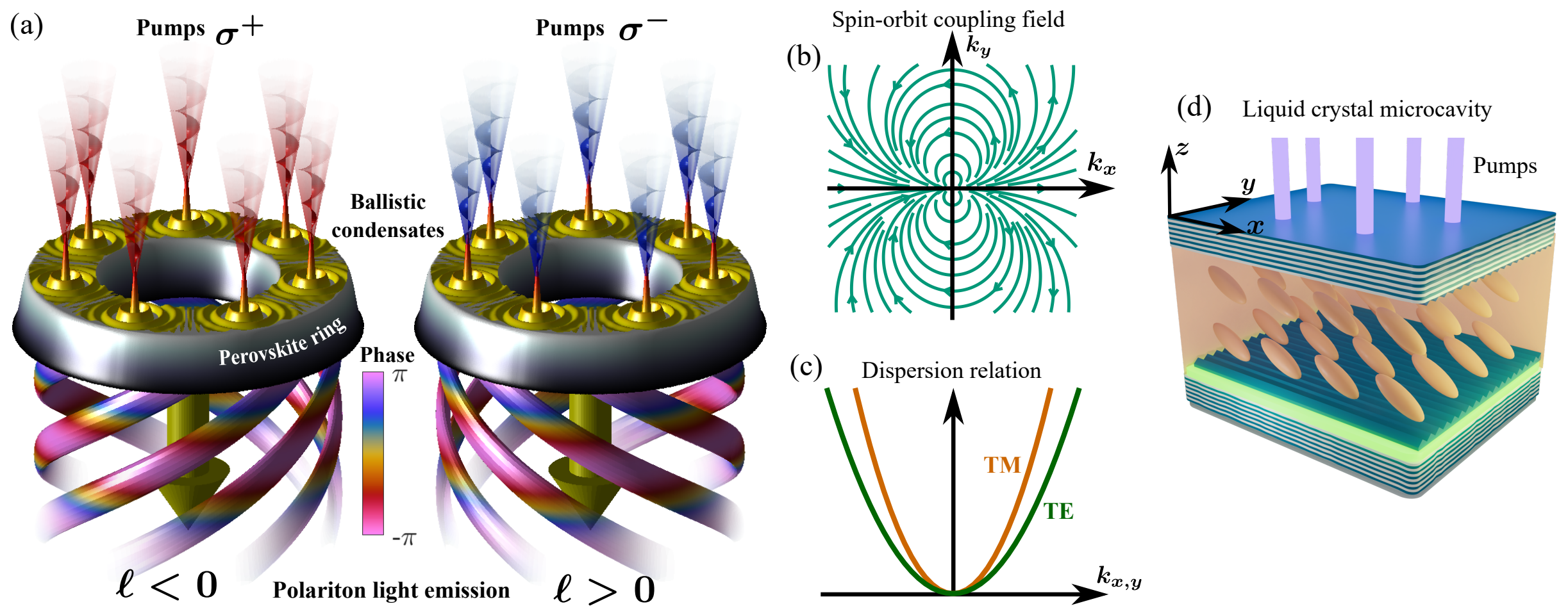

FIG. 1. (a) Scheme of the chiral polariton microlaser in transmission configuration. An odd number of equidistant nonresonant pump spots, structured using e.g. a spatial light modulator, are focused along the circumference of a semiconducting ring (e.g. perovskite crystal). The spots excite ballistic polariton condensates (yellow surfaces) which spontaneously obtain a circulating mass current with orbital angular momentum (OAM) $\ell$ locked with the spin angular momentum (SAM) $\sigma^\pm$ of the pump laser. (b) Green arrow map showing the SOC field texture $\mathbf{\Omega} \propto [\cos(2\theta), \sin(2\theta)]$ in the cavity momentum plane $(k_x, k_y)$ and (c) corresponding schematic TE-TM split lower polariton in-plane dispersion from Eq. (2). (d) Example schematic birefringent liquid crystal cavity made of distributed Bragg reflectors that can also be implemented in our multispot excitation scheme to achieve high values of SOC.

## II. RESULTS

### A. Spinor cavity polaritons

When excitons couple coherently with cavity photons they form two branches known as the "upper" and "lower" polariton branches. Condensation of polaritons preferentially occurs in the lower branch and therefore we will exclude the upper branch from our considerations. Similar to lasing, polariton condensation is a driven-dissipative and stimulated process where, in principle, multiple macroscopically coherent stable states can coexist and emit light [16]. An incoherent optical pump that is detuned far from the polariton level photoexcites hot excitons that relax down the polariton branch and provide gain to the condensate. In a "winner-takes-all" scenario, the state of lowest losses crosses threshold first and condenses while suppressing all other states. Importantly, due to the strong spin-dependent interactions of polaritons, owing dominantly to the exciton exchange term [54, 55], a $\sigma^\pm$ circularly polarized nonresonant excitation leads to a co-polarized condensate because of optical orientated bright $s = \pm 1$ excitons [56, 57]. This explicit optical breaking of chiral symmetry for polaritons can manifest in specific low-threshold OAM-carrying states by virtue of effective cavity-photon SOC [58]. That is, pumped polariton SAM can convert into OAM and be detected as optical vorticity in the emission [59]. In planar microcavities, this SOC comes from the reduced dimensionality which splits the TE and TM polarization modes as a function of the incidence angle $\sin(\vartheta_\text{in}) = ck/\omega$ where $|\mathbf{k}| = k$ and $\mathbf{k} \equiv (k_x, k_y)^\text{T}$ is the in-plane momentum of polaritons (photons) [60]. This momentum dependent splitting can be represented as an effective SOC field of strength $\beta$ [58] [see Fig. 1(b)],

$$\mathbf{\Omega} = \beta \begin{pmatrix} k_x^2 - k_y^2 \\ 2k_x k_y \\ 0 \end{pmatrix}. \quad (1)$$

that couples with the polariton pseudospin $\mathbf{S} = \Psi^\dagger \hat{\boldsymbol{\sigma}} \Psi$ causing it to processes. Here, $\hat{\boldsymbol{\sigma}}$ is the Pauli matrix vector.

At sufficiently long wavelengths in the lower polariton branch, we can work within the parabolic approximation and the Hamiltonian for SOC lower-branch spinor polaritons $\Psi = (\psi_+, \psi_-)^\text{T}$ in the absence of pumping is written,

$$\hat{H} = \frac{\hbar^2 k^2}{2m} + \beta k^2 \begin{pmatrix} 0 & e^{-2i\theta} \\ e^{2i\theta} & 0 \end{pmatrix} - i\frac{\hbar\gamma}{2}, \qquad (2)$$

where $m$ is the polariton effective mass, $\gamma$ its inverse lifetime, and $\theta$ is the angle of the polariton momentum in the cavity plane. Notably, the SOC operator represents an effective in-plane magnetic field which performs a double angle rotation in momentum space [see Fig. 1(b) and 1(c)] giving rise to momentum-dependent polariton spin precession [34, 58, 61, 62]. The strength of TE-TM splitting $\beta$ depends on the spectral position of the cavity photon mode with respect to the stop-band of the cavity [60]. For polaritons, this value can be further adjusted by changing the exciton-photon composition by varying their detuning which changes their Hopfield fractions [11]. Recent extremely birefringent liquid crystal microcavities [see Fig. 1(d)] have reported up to $\tilde{\beta} = m\beta/\hbar^2 \approx 0.1$ [63] for polaritons.

### B. Ring with uniform gain and losses

We start by considering SOC spinor polariton angular harmonics $e^{il\theta}$ in a uniformly pumped quantum ring of radius $R$ and OAM $l \in \mathbb{Z}$. Such polariton rings can be constructed anywhere from etched GaAs based materials [39] to patterned Ruddlesden–Popper perovskite layers [64] and template-assisted growth of single-crystal lead halide perovskites [? ], as a few examples. Our results are general to a planar cavity with no radial confinement but, for simplicity, we will focus on a quantum ring with one spatial variable $\mathbf{r} = R\hat{\boldsymbol{\theta}}$. We will neglect the optical Zeeman effect [20] which splits the energies of the polariton spins under circularly polarized pumping and instead focus on split loss-rates (linewidths) $\gamma_\pm = \gamma \pm \Gamma$. Our starting point is therefore more appropriate for a photon laser operating in the weak coupling regime where pump-induced blueshifts are negligible. We will address the role of spin-dependent polariton blueshifts in the next section. For brevity, we set $\hbar = mR^2 = 1$ and work with dimensionless units. The non-Hermitian SOC polariton "Hamiltonian" is written:

$$\hat{H} = \begin{pmatrix} -\frac{\partial_\theta^2}{2} - i(\gamma - \Gamma) & \tilde{\beta} e^{-2i\theta}\left(\partial_\theta^2 - 2i\partial_\theta\right) \\ \tilde{\beta} e^{2i\theta}\left(\partial_\theta^2 + 2i\partial_\theta\right) & -\frac{\partial_\theta^2}{2} - i(\gamma + \Gamma) \end{pmatrix}. \qquad (3)$$

Here, $\partial_\theta$ is the partial derivative along the ring angle $\theta$ and the strength of the SOC is denoted by the dimensionless parameter $\tilde{\beta}$. The anti-Hermitian field $i\Gamma\hat{\sigma}_z$ represents a "loss-imbalance" between the polariton spins coming from circularly polarized pumping which spin polarizes the gain medium through the optical orientation effect. From here on, we will set $\gamma = 0$ without loss of generality and, for simplicity, neglect energy-dependent losses and scattering from the reservoir into the polariton modes (i.e., broad gain-bandwidth assumption).

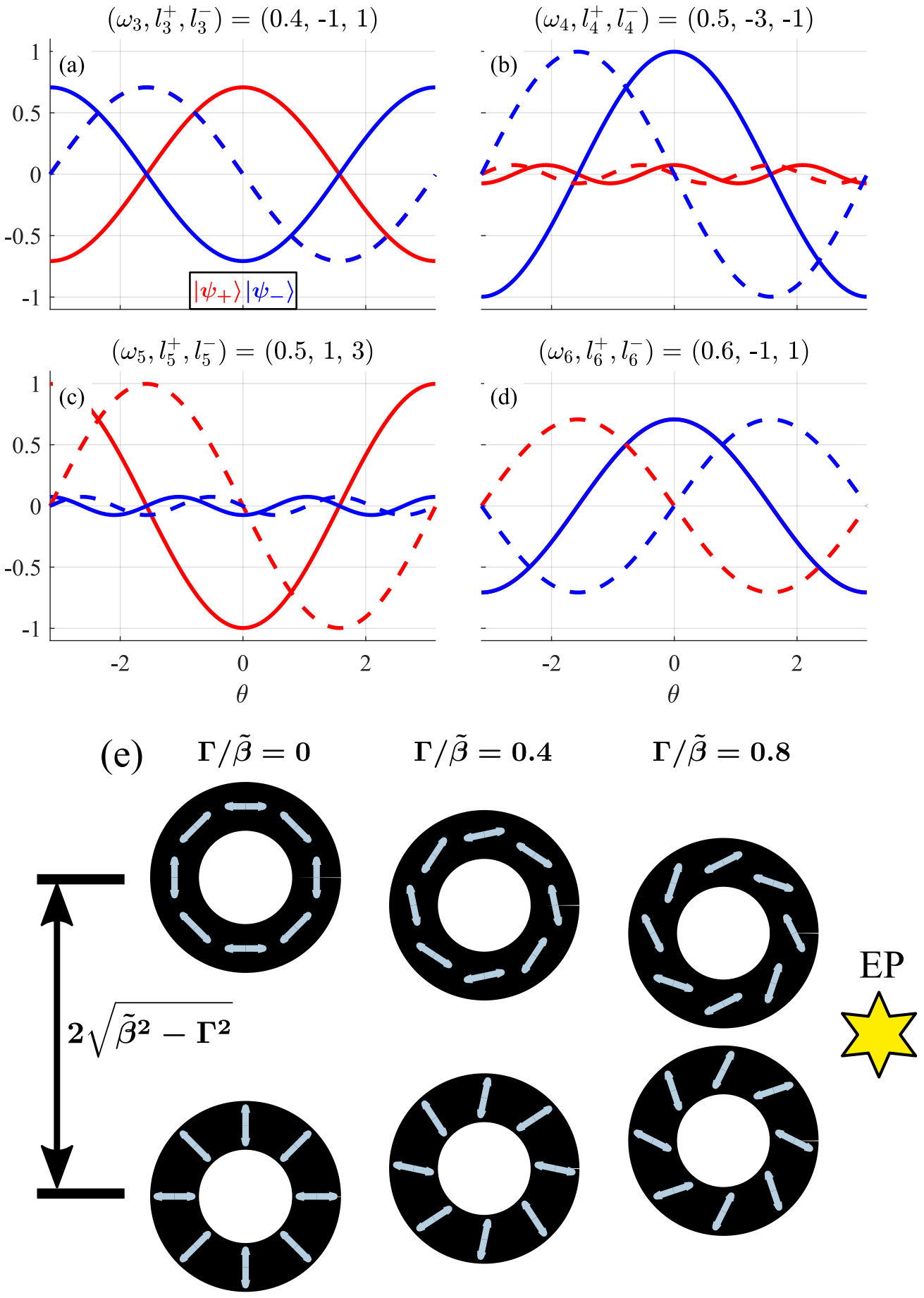

FIG. 2. (a-d) First four excited eigenstates of the polariton quantum ring with SOC $\tilde{\beta} = 0.1$ and $\Gamma = 0$. (e) Effect of the non-Hermitian field on the linearly polarized eigenstate corresponding to panels (a) and (d). Double-headed blue arrows denote the linear polarization along the ring. As the strength of $\Gamma$ increases the two states tilt until they coalesce into an exceptional point (EP).

Since the system possesses continuous rotational symmetry we can define an OAM operator $\hat{L}_z = -i\partial_\theta$ in the standard fashion along the out-of-plane $z$-direction. In the trivial case of $\tilde{\beta} = 0$, $\hat{H}$ and $\hat{L}_z$ commute and share spinor-valued eigenstates $\Psi_l^\pm = e^{il\theta}|s = \pm 1\rangle/\sqrt{2\pi}$ with eigenvalues $\epsilon_{\pm,l} = l^2/2 \pm i\Gamma$.

When $\tilde{\beta} \neq 0$ the off-diagonal SOC term couples spins differing by 2 quanta of OAM, conserving net angular momentum $J = l + s$. Here, SAM and OAM are no longer good quantum numbers and we will instead label our eigenvalues $\epsilon_n = \omega_n + i\upsilon_n$ and eigenstates $\Psi_n$ where $n = 1, 2, 3, \ldots$. In this sense, $\omega_n$ denotes the energy of the $n$th eigenstate whereas positive(negative) valued $\upsilon_n$ denote its growth(decay) in time. We also define the total OAM of each eigenstate,

$$l_{\rm tot} = \int_0^{2\pi} \Psi_n^\dagger (\mathbb{1}_{2\times 2} \otimes \hat{L}_z)\Psi_n \, d\theta. \quad (4)$$

Figures 2(a)-2(d) show the first four excited "$p$-states" of the quantum ring in the Hermitian case of $\Gamma = 0$ (the trivial twofold degenerate ground $s$-states $\Psi_{l=0}^{\pm}$ are not shown). The red and blue colors denote spin-up and spin-down components of the spinor wavefunction. Solid and dashed curves denote the real and imaginary parts of the wavefunction. Notably, Figs. 2(a) and 2(d) show $p$-states with zero net OAM and equal amounts of spin-up and spin-down projections $|\psi_\pm\rangle$ because the SOC couples $\Psi_{l=-1}^{+} \leftrightarrow \Psi_{l=+1}^{-}$ together [19–21]. These linearly polarized $p$-states are polarized parallel 2(a) and perpendicular 2(d) to the ring trajectory [see Fig. 2(e) far left] and are split by amount $\omega_6 - \omega_3 = 2\tilde{\beta}$. In between, are a pair of degenerate states in the $|J| = 2$ manifold [see Fig. 2(b) and 2(c)] that were crucial for the demonstration of a chiral photon [65] and polariton laser [20].

When the anti-Hermitian term $\Gamma$ is included the spectrum becomes complex-valued $\upsilon_n \neq 0$, as shown in Figs. 3(a) and 3(b). The red-blue color scale denotes the normalized spin polarization of the state averaged over the circle $s_{n,z} = \int \Psi_n^\dagger \hat{\sigma}_z \Psi_n d\theta / \int \Psi_n^\dagger \Psi_n d\theta$ (i.e., the degree of circular polarization of the emitted light). Here we have chosen $\Gamma > 0$ which means the red colored spin-up states have higher imaginary energy and therefore "live" longer (i.e., have higher gain). The gray dot in Figs. 3(a) and 3(c) at $\upsilon_n = 0$ corresponds to the orthogonal linearly polarized states discussed in Fig. 2(a) and 2(d). For small values of SOC the eigenvalues of these two linearly polarized states are given by $\epsilon_n = 1/2 \pm \sqrt{\tilde{\beta}^2 - \Gamma^2}$. The presence of the anti-Hermitian field $i\Gamma\hat{\sigma}_z$ tilts the linear polarization vector, so the states are no longer parallel and perpendicular with respect to the ring trajectory [see Fig. 2(e)]. For increasing $\Gamma$, the two linearly polarized eigenstates coalesce at the degeneracy point $\tilde{\beta} = \Gamma$ (yellow star) and obtain finite and opposite circular polarizations $|s_z| > 0$ denoted by black arrows in 3(b) and 3(d). Such a point is known as an *exceptional point* which has been subject of much discussion in optics [66] and polaritonics [67, 68]. We will revisit this phenomenon in a future study.

The chiral principle of the SOC system is captured in the skewed total OAM distributions of the imaginary energies seen in Fig. 3(c) and 3(d), illustrated with the colored horizontal arrows. Namely, high gain spin-up (red) states are shifted towards negative OAM states whereas large loss spin-down (blue) states are shifted towards positive OAM states. Therefore, if the system is pumped above threshold the dominant lasing modes belong spin-up polaritons biased towards clockwise rotating vortex states. If the loss-imbalance field $i\Gamma\hat{\sigma}_z$ is flipped (i.e., pump circular polarization flipped) then the dominant lasing modes will instead belong to spin-down polaritons in counterclockwise vortex states.

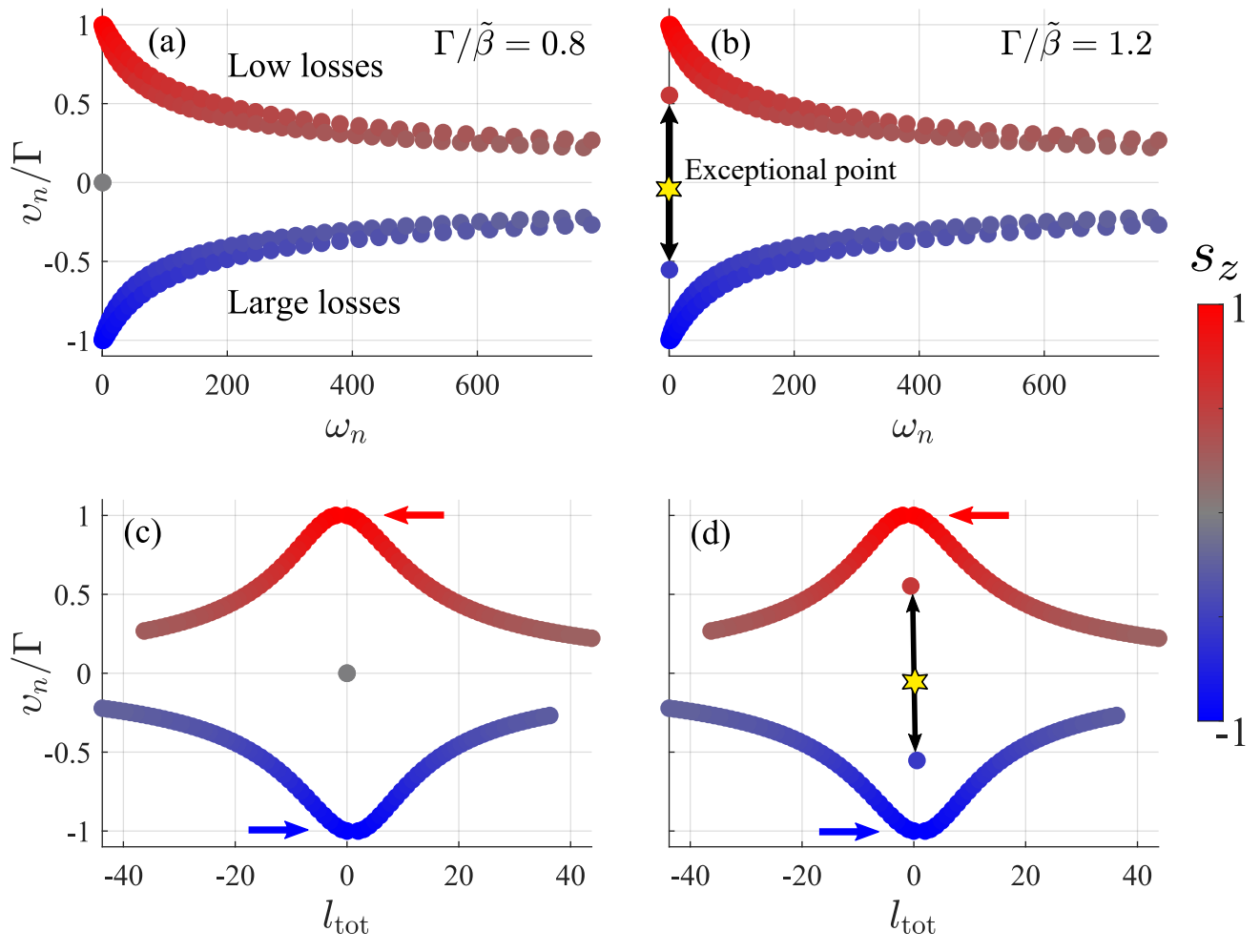

FIG. 3. (a,b) Complex eigenenergies of the polariton ring states (truncated to a basis of $|l| \leq 40$) for two different values of $\Gamma$. (c,d) Same set of eigenvalues but now plotted as function of $l_{\rm tot}$. The exceptional point is marked with a yellow star corresponding to two linearly polarized states coalescing and obtaining opposite $z$-spin projection (circular polarization) marked by black arrows.

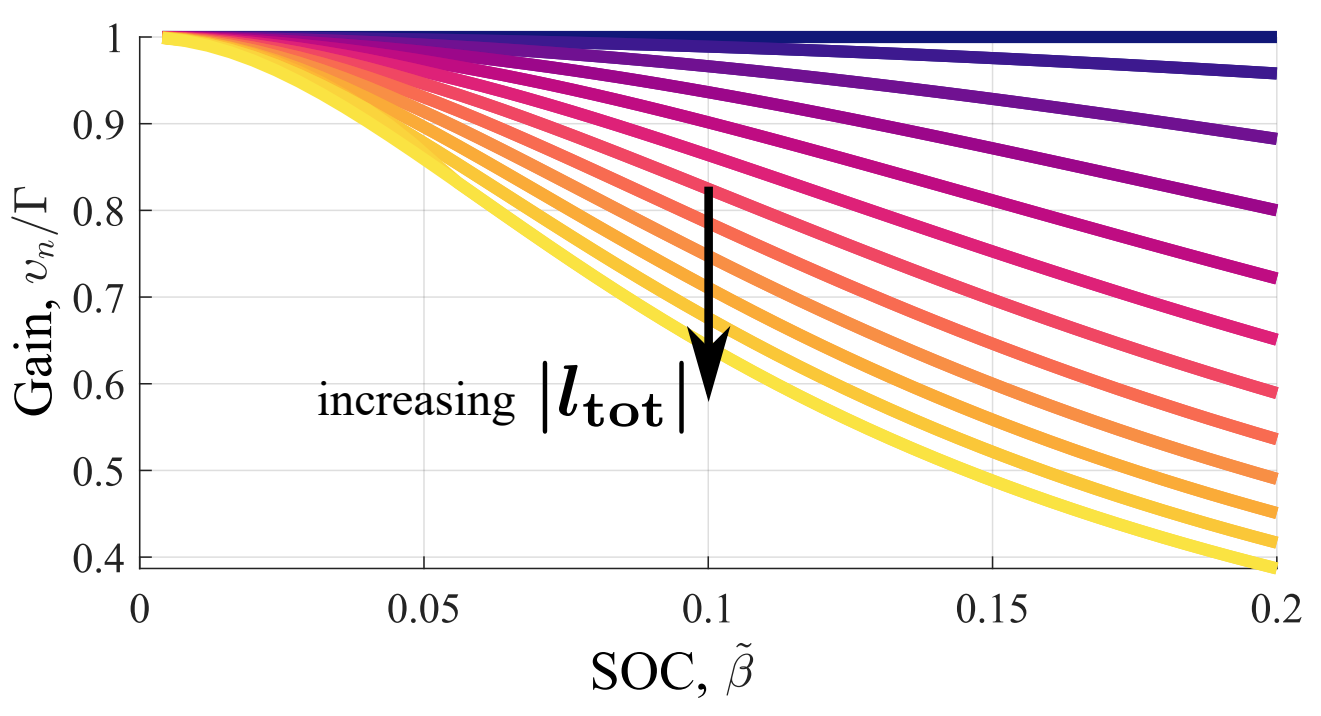

FIG. 4. Largest imaginary energies as a function of SOC strength in a unifromly pumped polariton quantum ring. The decreasing position of the curves represents an increasing amount of OAM that mixes the spins more efficiently, distributing gain and losses to reach $\upsilon_n \to 0$.

Figure 4 shows how the largest positive imaginary energies monotonically decrease as a function of SOC strength. The lower curves carry higher amounts of OAM corresponding to more equal mixtures of the spins, causing the imaginary energy to approach zero $\upsilon_n \to 0$. The highest horizontal line at $\upsilon_n/\Gamma = 1$ corresponds to pure spin-up states $\Psi_{l=0}^{+}$ and $\Psi_{l=-2}^{+}$ with eigenvalues $\epsilon = i\Gamma$ and $\epsilon = 2 + i\Gamma$, respectively. These specific states cannot couple to the lossy spin-down component $\Psi_l^{-}$ because of lack of momentum in either spin component. The eigenvalues therefore don't depend on the strength of SOC and form a hard upper bound for all other imaginary energies. This is the key reason why we have neglected the optical Zeeman effect in this section because it cannot change this upper bound. The next-largest imaginary energy curve belongs to a pair of states with $(l^+, l^-) = (-3, -1)$ and $(1, 3)$. These states have eigenvalues $\epsilon = 5/2 \pm \sqrt{(2 \pm i\Gamma)^2 + (3\tilde{\beta})^2}$ and therefore become lossier with increasing SOC since the up-spins are now coupled with the lossy down-spins. The same trend is true for all following states. This upper bound forms an obstacle in design of a flexible chiral polariton lasing device which we overcome in the next section.

### C. Ring with structured gain and losses

The results of previous section illustrate the chiral response of the SOC polariton ring when the gain medium is uniformly spin polarized and how SOC redistributes the gain between the angular harmonics. However, this response is limited from a practical point of view. **i**) First, the eigenstates $\Psi^+_{l=-2}$ and $\Psi^+_{l=0}$ have equal maximum gain and therefore, when pumped above threshold, the system would lase with equal probability in either state. The winner would be randomly chosen by spontaneous processes around threshold. Such ambiguity is detrimental for a system that should operate deterministically. **ii**) Second, only the $|l| = 2$ states carry finite OAM that are locked to the pump polarization $\Psi^{\pm}_{l=\mp 2} \leftrightarrow \sigma^{\pm}$ and cannot be adjusted. That is, other OAM states will always have lower gain. This is problematic from the point of view of designing a system where variable OAM needs to be selected by tuning the system parameters, important for chiral optical technologies. **iii**) Third, the contrast between the largest and next-largest imaginary energy is weak, only about $\approx 1\%$ even for generous values of SOC $\tilde{\beta} = 0.1$ [compare top two curves in Fig. 3]. Sufficiently strong environmental fluctuations can therefore cause accidental lasing in undesirable neighboring state manifolds.

The above challenges can all be addressed by replacing continuous rotational symmetry instead with discrete rotational symmetry of odd-numbered order (i.e., triangle, pentagon, heptagon, etc.). This can be achieved by spatially structuring the nonresonant pump profile so that it satisfies $P(\theta) = P(\theta + 2\pi/N)$. The system then belongs to the point-group rotational symmetry class $\mathcal{C}_N$ of order $N$ where $N$ is an odd number, thus breaking in-plane inversion symmetry. This strategy is motivated by the ability of polaritons to form spontaneous currents in periodic potential-gain landscapes [69] and the recent demonstration of geometrically frustrated ballistic polariton condensates spontaneously forming giant discrete vortices [47]. To this end, we define a pump profile built using equidistant Gaussian pump spots along the ring circumference [see example schematic Fig. 1 for $N = 7$],

$$P_{\pm}(\theta) = P_0 \sum_{n=1}^{N} e^{-(\theta - 2\pi n/N)^2/2w^2} \cos^2\left(\Theta \mp \frac{\pi}{4}\right), \quad (5)$$

where $P_{\pm}(\theta)$ refers to right-hand and left-hand circularly polarized photons, $P_0$ is the power density, and $\Theta$ defines the pump ellipticity. The width of each spot $w$ is taken to be smaller than their intersite distance, i.e. $2\pi/N \gg w$, which facilitates the phenomena of ballistic polariton condensation [70, 71]. An example pump profile is shown within the ring unit cell in Fig. 5(a). Because of their much larger mass the photoexcited reservoir of excitons is co-localized with the pump spots, forming an effective optically generated potential landscape that repels polaritons due to the strong interactions [72, 73]. The Hamiltonian therefore obtains a spatially dependent term and is now written,

$$\hat{H} = \begin{pmatrix} -\frac{\partial_\theta^2}{2} + (1 + i\Gamma)P_+(\theta) & \tilde{\beta} e^{-2i\theta}\left(\partial_\theta^2 - 2i\partial_\theta\right) \\ \tilde{\beta} e^{2i\theta}\left(\partial_\theta^2 + 2i\partial_\theta\right) & -\frac{\partial_\theta^2}{2} + (1 + i\Gamma)P_-(\theta) \end{pmatrix}. \quad (6)$$

Here, $(1 + i\Gamma)P_{\pm}$ represents the amount of blueshift and gain felt by each polariton spin component due to the elliptically polarized pump [74]. The energy difference $\Delta = P_+ - P_-$ constitutes an interaction-induced optical Zeeman splitting [20]. Different from the uniformly pumped quantum ring in the previous section, this blueshift plays now an important role in determining the optimal lasing state when the pumping is spatially dependent. Note, the parameter $\Gamma$ still holds the same meaning as a "gain imbalance" parameter when $P_+ \neq P_-$.

Because of the discrete rotational symmetry, the standard OAM defined by the operator $\hat{L}_z$ is no longer a conserved quantity and instead becomes defined by angular Bloch momentum $\ell \in \mathbb{Z}$, or quasimomentum of the Bloch wavefunction [75],

$$\Psi^{\sigma}_{\ell,\nu} = e^{i\ell\theta} u^{\sigma}_{\ell,\nu}(\theta)|\sigma\rangle. \quad (7)$$

whose envelope satisfies $u^{\sigma}_{\ell,\nu}(\theta) = u^{\sigma}_{\ell,\nu}(\theta + 2\pi/N)$. For simplicity, we will still refer to the quasimomentum $\ell$ as OAM. Here, the index $\nu$ denotes the band index and, restricting ourselves within the first Brillouin zone [see Fig. 5(b)], the permitted values of quasimomentum for odd number of pump spots are $|\ell| \leq \frac{N-1}{2}$ in order to satisfy the symmetry of the system. Note that if $N$ is an even integer the inequality becomes $|\ell| \leq \frac{N}{2}$. As a shorthand notation, we will refer to the Bloch OAM in its respective spin component as $\ell^{\pm}$. Figure 5(b) shows the calculated lowest real energy bands in the first Brillouin zone of the ring under elliptically polarized pumping without SOC $\tilde{\beta} = 0$. In this case the Bloch eigenstates are pure spin-up (red) or spin-down (blue) states split by a characteristic effective "Zeeman" energy $\Delta \propto P_+ - P_-$. The red and blue markers denote the allowed quasimomenta for $N = 7$ heptagonal pump geometry. The size of the markers indicates the amount of gain $\upsilon_n$ in the state. The thin curves connecting the discrete quasimomenta denote Bloch bands in the continuum limit (infinite lattice), as a reference. When SOC is included, spin-mixing occurs between Bloch states differing by $\ell^- - \ell^+ = 2$, schematically shown by the example green double-headed arrows.

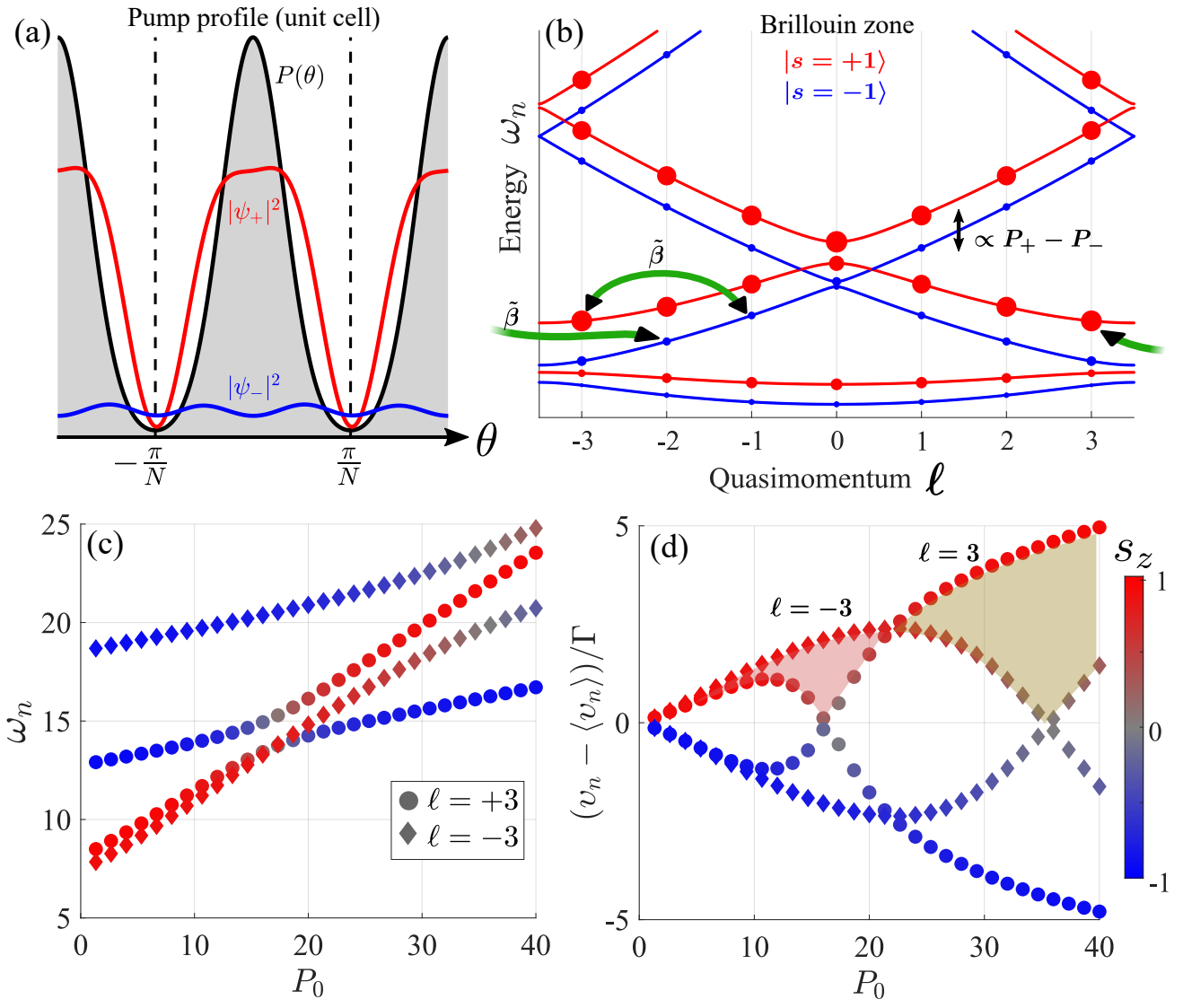

FIG. 5. (a) Example pump (potential) profile within the unit cell of the ring and corresponding highest gain Bloch solution $|\psi_\pm|^2$. (b) Calculated energy structure in the first Brillouin zone of the ring without SOC. Red and blue thin curves denote the bands in the continuum limit (infinite lattice) whereas the dots denote allowed quasimomenta for $N = 7$. The size of the dots indicates the amount of gain $\upsilon_n$ in that state. Green double headed arrows show schematically SOC channels. (c) Real and (d) imaginary parts of four high gain Bloch eigenvalues as function of pump power. The $\langle \upsilon_n \rangle$ denotes their average imaginary part. Other parameters: $N = 7$, $\tilde{\beta} = \Gamma = 0.1$, $w = N^{-1}$, $\sin(2\Theta) = 0.5$.

Amplified polariton Bloch states belong to a certain band index $\nu$ and quasimomenta $\ell$ which optimizes constructive interference over the pumped region [69]. These states usually reside in the high symmetry points of the Brillouin zone and possess the lowest condensation threshold as observed in past experiments on optical polariton lattices [76]. For example, compare the different sizes of the red markers in Fig. 5(b) between $|\ell| = 3$ and $\ell = 0$ and between bands. When $N$ is even numbered on the ring, these points are the $\Gamma$-point ($\ell = 0$) and the $M$-point ($|\ell| = N/2$) which don't carry any net OAM. However, when $N$ is odd numbered, the edge of the ring's Brillouin zone is forbidden by symmetry, and the nearest available state becomes $|\ell| = (N-1)/2$ which does carry finite OAM. By tuning the pump parameters, this frustrated "edge" state can be engineered to have the highest gain [47] thus resulting in persistent circulating polariton currents or *giant discrete vortices.* These are also known as "twisted states" in arrays of nonlinear oscillators [77].

This phenomena becomes much richer in the presence of SOC which facilitates tunable chiral polariton lasing. To demonstrate this, we set the number of pump spots to $N = 7$, the ellipticity to $\sin(2\Theta) = (P_+ - P_-)/(P_+ + P_-) = 0.5$, and $\tilde{\beta} = \Gamma = 0.1$ and numerically diagonalize Eq. (6). In order to not overcrowd the data, we plot only a few eigenvalues with the highest imaginary part belonging to the edges of the Brillouin zone as filled diamonds ($\ell = -3$) and circles ($\ell = 3$) in Figs. 5(c) and 5(d) as a function of pump power $P_0$. For low powers the highest imaginary part belongs to the $\ell = -3$ state (pink shaded area) which suddenly reverses to $\ell = 3$ (yellow shaded area) for larger powers. This transition is accompanied by inter-band mixing of the spins which can be clearly observed as an avoided crossing in the real energies in Fig. 5(c). Indeed, as the power increases, spins belonging to different bands shift into one-another and couple resonantly. The complicated redistribution of losses allows other Bloch states to overtake the initial winner. The non-monotonic rise and fall of the imaginary eigenvalues underpins the tunability of structured pumping which is in sharp contrast to the bounded eigenvalues found in the uniformly pumped quantum ring in Fig. 4.

Amazingly, multiple different regions of winning lasing modes carrying OAM exist in the parameter space of the system, underlining its tunability. Figure 6(a) shows the total OAM of the highest gain Bloch eigenstate as a function of both pump power and ellipticity. A rich phase diagram appears, revealing regions of definite OAM locked with the polarization of the pump. That is, if we reverse the sign of $\Theta$ the sign of the OAM also reverses. At each pixel, the highest gain eigenstate was dominantly co-polarized with the pump, with an average value of $s_z \approx 0.86 \pm 0.08$ within the investigated window. This simply underlines the fact that the SAM of the pump is mostly transferred to the spin of the polariton. Figure 6(b) shows the contrast between the largest and the next-largest imaginary energy $= (\upsilon_1 - \upsilon_2)/\upsilon_1$. By simply structuring the pump we obtain now a much higher gain contrast $\approx 14\%$ as compared to the $\approx 1\%$ in the uniformly pumped ring. Importantly, switching between OAM states is possible by adjusting either the power or the polarization of the pump, which can be done using an acousto-optic modulator at MHz rates or photoelastic modulator at hundreds of KHz, respectively, for typical excitation wavelengths of e.g. lead halide perovskites at $\sim 400$ nm. Importantly, the condensate vortex switching time is expected to be in the order of picoseconds determined by the typical cavity polariton timescales.

In order to verify that these states grow the fastest and quench all other modes one should employ a mean-field model based on the generalized Gross-Pitaevskii equation (GPE) coupled to an incoherent exciton reservoir as is the usual practice when dealing with exciton-polariton condensate dynamics [11]. We will instead start by using a simpler 1D complex Ginzburg-Landau equation (cGLE) describing the spinor-valued polariton order parameter $\Psi(\theta) = [\psi_+(\theta), \psi_-(\theta)]^{\text{T}}$ on the ring,

$$i\frac{\partial \Psi}{\partial t} = \hat{H}\Psi - i\begin{pmatrix} |\psi_+|^2 & 0 \\ 0 & |\psi_-|^2 \end{pmatrix}\Psi. \tag{8}$$

Here, $\hat{H}$ is given by Eq. (6) and the role of the nonlinear term is to make the system convergent (i.e., gain clamping). The cGLE can be obtained from the GPE by assuming that the exciton reservoir relaxes fast compared to the condensate density fluctuations, and that blueshift from polariton-polariton interactions is much weaker compared to other energy scales in the system. We numerically integrate the above equation starting from random white-noise initial conditions using parameters corresponding to the four cases indicated by the red markers in Fig. 6(a).

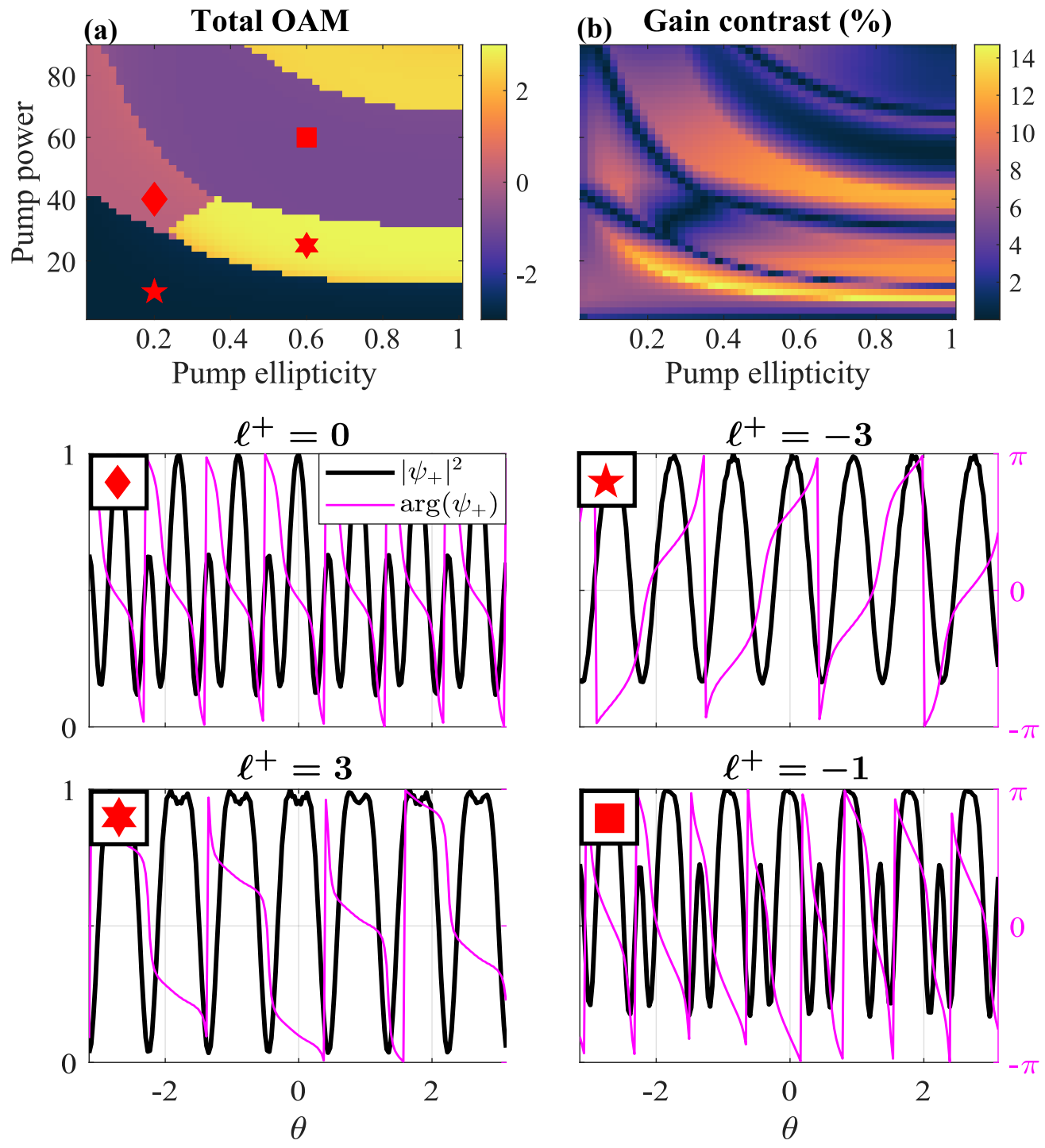

FIG. 6. (a) Total angular momentum $\ell_{\text{tot}}$ of the highest gain state for $N = 7$ spots as a function of pump power $P_0$ and ellipticity $\sin(2\Theta)$. (b) Corresponding difference between the highest and the next-highest imaginary energy $(\upsilon_1 - \upsilon_2)/\upsilon_1$. (Bottom panels) Dominant fixed-point solutions obtained from cGLE simulations corresponding to pump parameters indicated by red markers in (a). Parameters: $\tilde{\beta} = \Gamma = 0.1$ and $w = 1/N$.

The density (black lines) and phase (magenta lines) of the dominantly pumped $\psi_+$ component of the resulting fixed-point solutions are shown in the bottom panels of Fig. 6, underlining that the polariton order parameter rises quickly to the highest gain solution and quenches all others. The $\psi_-$ component is not shown since it is very weak. Notably, the pentagram, star, and square markers represent solutions with a persistent current of OAM $\ell^+ = -3, 3, -1$. These are giant discrete polariton vortices of variable OAM depending on the pump parameters. The stability of these states is verified by repeating the simulations over 100 different initial conditions which all converged to the correct solution. We stress that if the ellipticity of the pump is flipped then the OAM of the pumped component flips as well, underlining OAM-SAM locking.

| | |
|---|---|
| $\hbar\gamma$ | 0.1 meV |
| $m$ | 0.3 meV ps$^2$ $\mu$m$^{-2}$ |
| $\lvert X\rvert^2$ | 0.5 |
| $g_0$ | 0.01 meV $\mu$m$^2$ |
| $2\sqrt{2\ln 2}w$ | 3 $\mu$m |
| $\hbar\Gamma_R$ | $20g_0$ |
| $2m\beta/\hbar^2$ | 0.1 |
| $\eta\Gamma_R$ | 2 |

Table I. Parameters of the 2D GPE simulations. Other parameters are: $\Gamma_X = \Gamma_s = \gamma$.

### D. Planar cavity: 2DGPE simulations

Up until now we have considered the eigenmodes and dynamics of polaritons on a 1D quantum ring. In this section we extend our analysis to a more general 2D planar cavity without radial confinement. Although the presence of a quantum ring can help lower the threshold of the system and guide polariton circulation, it is not strictly necessary since discrete rotational symmetry is still imposed by the pumps focused onto the cavity plane [47]. The dynamics of the planar polariton condensate, and its stable solutions, are well described using the spinor 2D GPE coupled to an exciton reservoir [11]. Here, we will use physical units and parameters guided from previous experiments given in Table I (see Supplemental Material for further information on parameter choices).

$$i\hbar\frac{\partial\psi_\pm}{\partial t} = \left[-\frac{\hbar^2\nabla^2}{2m} + g|\psi_\pm|^2 + g_R\left(n_\pm + \eta P_\pm\right) + i\hbar\frac{(\Gamma_R n_\pm - \gamma)}{2}\right]\psi_\pm + \beta(i\partial_x \pm \partial_y)^2\psi_\mp \tag{9}$$

$$\frac{\partial n_\pm}{\partial t} = -(\Gamma_X + \Gamma_R|\psi_\pm|^2)n_\pm + \Gamma_s(n_\mp - n_\pm) + P_\pm(\mathbf{r}) \tag{10}$$

Here, $\nabla^2 = \partial_x^2 + \partial_y^2$ is the 2D Laplacian, $g = g_0|X|^4$ and $g_R = 2g_0|X|^2$ are the same spin polariton-polariton and polariton-exciton interaction strengths, $|X|^2$ is the polariton's excitonic Hopfield fraction, $\Gamma_R$ is the spin-dependent stimulated scattering rate of polaritons from the active reservoir to the condensed state, and $\gamma$ is the average polariton decay rate. $\Gamma_X$ is the decay rate of active bottleneck excitons, $\eta P_\pm(\mathbf{r})$ denotes additional blueshift coming from a background of optically inactive dark excitons, and $\Gamma_s$ describes spin-relaxation within the exciton reservoirs present in GaAs quantum wells [78, 79] and in organic-inorganic halide perovskites [80]. The pump pattern is made up of Gaussians that lie on the vertices of an odd numbered polygon [see inset in Fig. 7],

$$P_\pm(\mathbf{r}) = P_0\sum_{n=1}^{N} e^{-|\mathbf{r}-\mathbf{r}_n|^2/2w^2}\cos\left(\Theta \mp \frac{\pi}{4}\right). \tag{11}$$

Here, $\mathbf{r}_n = R[\cos(\theta_n), \sin(\theta_n)]$ and $\theta_n = 2\pi n/N$ are the coordinates of the pumps lying at the vertices of a polygon of radius $R$. Each pump spot generates a ballistic condensate with a radially expanding wavefunction of high kinetic energy polaritons that travel to distant neighbours, coupling together the condensates [71, 81]. Depending on the separation distance between two condensates, they can synchronize either in-phase, anti-phase, or with fractional $\Delta\phi = 2\pi n/N$ phase slips depending on the geometry [47]. The pump power density is chosen around $P_0 \approx 1.2 P_{\text{th}}$ where $P_{\text{th}}$ is the condensation threshold power defined by the condition where the largest imaginary part of the linearized 2DGPE energies (9) becomes positive. Working close to threshold means that the lowest-loss OAM solution emerges first from random white-noise initial conditions and quenches other competing modes. In real units, the typical threshold power (fluence) in pulsed experiments in both GaAs [82] and lead halide perovskites [83–85] microcavities is in the order of $\sim \mu$J cm$^{-2}$, lower than typical powers in optically pumped III-V photon lasers.

The radial degree of freedom introduces transverse losses which now play a role in determining the lowest threshold winner and stabilizing the pumped condensate. Figure 7 shows a realization map of the total OAM obtained from a condensate pumped in a heptagonal geometry ($N = 7$) as a function of pump polygon radius $R$ and ellipticity $\sin(2\Theta)$. The inset shows an example pump profile. Each pixel represents a convergent simulation from a random initial condition, mapping out attractors in the system phase space with definite OAM (labeled with $\ell^{\pm}$). These results evidence that a chiral ballistic polariton laser can operate on an open cavity plane without any pre-fabricated quantum ring potential. Interestingly, for larger values of ellipticity one can find a speckled region where the condensate converges to either of the two dominant and counter-rotating vortex solutions, implying bistability. The slight color variation in some of the pixels corresponds to simulations that did not fully converge. Vortex solutions for more types of pump polygons are shown in the Supplemental Material.

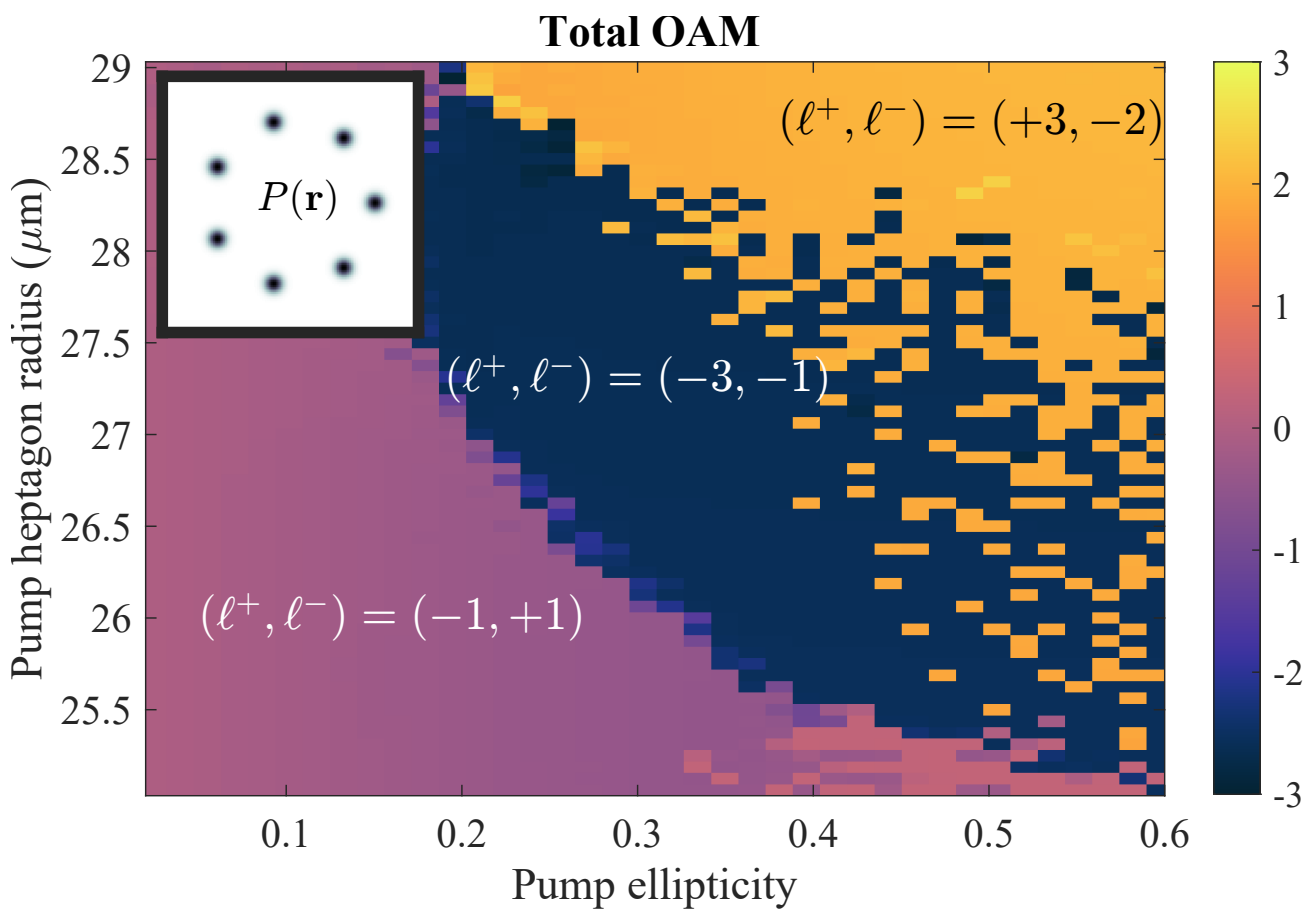

FIG. 7. Total OAM from the condensate final state obtained from 2DGPE simulations as a function of pump polygon radius $R$ and ellipticity $\sin(2\Theta)$. Each pixel corresponds to a random realization of initial conditions (i.e., Monte Carlo trials) numerically integrated for a sufficiently long time. Three regimes are marked by their respective values of $(\ell^+, \ell^-)$. The inset shows an example pump profile. The power density $P_0$ is fixed and given in Table I.

Figure 8 shows the spatial profiles of the three types of condensate solutions found from Fig. 7. Here, the ballistic nature of the condensates is vividly captured in their expansion resulting in rich spiral shaped interference due to the OAM being carried by the spins. The two leftmost columns show the spatial density of each spin component. The weaker spin-down component $\psi_-$ is enhanced to more clearly see its profile. The second-from-right column shows the $s_z$ parameter, i.e. the degree of circular polarization of emitted light, displaying rich spin structure with multiple brightly polarized regions. Many of these regions correspond to a polarization-singular textures known as C-points where $|s_z| = 1$ and V-points where $|\psi_+|^2 + |\psi_-|^2 = 0$ with a winding linear polarization vector [52]. The singular behaviour of these points is more vividly seen in the Stokes phase of the condensate defined as $\phi_{xy} = \arg(s_x + i s_y)$ where $s_{x,y} = \int \Psi_n^\dagger \hat{\sigma}_{x,y} \Psi_n d\mathbf{r} / \int \Psi_n^\dagger \Psi_n d\mathbf{r}$. More specifically, the Stokes phase describes the angle of the linear polarization vector of the emitted light. In the rightmost column we show the Stokes phase zoomed around the origin indicated by the black dashed square in the neighboring column. Multiple regions of winding linear polarization result in Stokes phase singularities of varying topological charges, underlining the complex interference and phase-locking between the ballistic condensates. The composition of the singularities depends on the specific angular Bloch modes, for example, V-points of charge $\ell^+ - \ell^- = -2$ can be located at the origin in top and middle rows in Fig. 8 and a charge $\ell^+ - \ell_- = 5$ in the bottom row.

## III. DISCUSSION AND CONCLUSIONS

We have proposed an all-optically reconfigurable discrete chiral vortex microlaser based on ballistic polariton condensates that promises a wider range of polariton-optical vortex tunability than previous proposals and realizations. Our system is based on three important ingredients: i) the optical Zeeman effect, or spin-dependent polariton interactions; ii) effective spin-orbit coupling coming from inherent TE-TM splitting of the photonic cavity modes; and iii) breaking of continuous-rotational symmetry and inversion symmetry through structured pumping in odd-vertex polygons. The first two ingredients have already been exploited to demonstrate chiral polariton lasing in a single micropillar [20].

Our work juxtaposes the important effects of continuous versus discrete rotational symmetries on the spectrum of polariton modes carrying OAM and how the latter can amplify chiral response with multiple tunable parameters. By exploiting the phenomena of ballistic condensation we can achieve high charge vortex charge lasing with OAM locked to the pump SAM. Different orders of vortex charges can be selected through reconfigurable all-optical adjustment of the pump parameters such as power $P_0$, polarization $\Theta$, number of pump spots (vertices) $N$, polygon radius $R$, and spot width $w$. Such a form of flexibility goes beyond previous works on controlling polariton vorticity using explicit chiral pumping patterns [25], temporally modulating the pump to "stir up" vorticity [29, 30], or by direct phase transfer [86]. Our work has clear advantages in breaking chiral symmetry for polaritons over real magnetic fields [27] due to the small exciton $g$-factor in typical semiconductors.

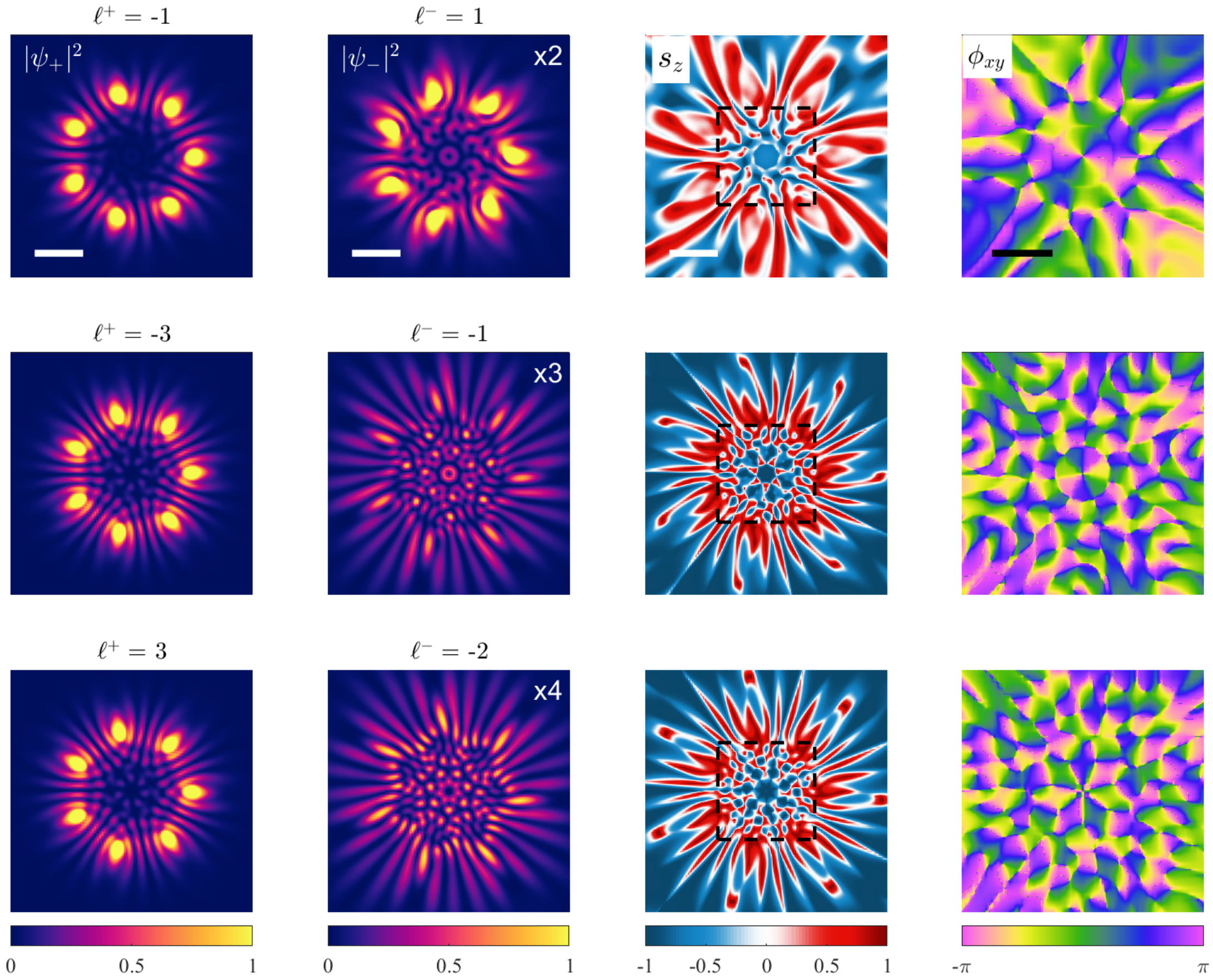

FIG. 8. Steady state giant vortex solutions obtained from spinor 2D GPE simulations (9). (Left to right) Condensate spin-up density $|\psi_+|^2$, spin-down density $|\psi_-|^2$, normalized degree of circular polarization $s_z$, and Stokes phase $\phi_{xy} = \arg(s_x + is_y)$. (Top to bottom) Different complexes of OAM corresponding to the regimes located in Fig. 7. Because the spin-down density is less populated for positive ellipticities we multiply the data with a scalar (upper right corner) to increase visibility in the plots. White scale bars are 20 $\mu$m and black 10 $\mu$m.

In our study, we have assumed a broad gain-bandwidth system by neglecting any explicit energy dependence of polariton losses $\gamma$ and scattering rates $\Gamma_R$ from the exciton reservoir into the condensate. However, it is known that the exact energies of polariton levels do play a role in selection of lasing states [20, 65]. We expect that this additional mechanism can complement our ballistic device by separating the high charge OAM lasing state even further from neighboring states.

Our study does not showcase an exhaustive list of available OAM carrying polariton states. Notably, we have restricted our analysis to mostly $N = 7$ discrete rotational geometries that permit orbital quasimomentum up to $|\ell| = 3$. Higher order $N$-polygonal pump patterns will give access to even higher values of quasimomenta bounded by $|\ell| \leq (N-1)/2$ and thus faster rotating currents (see Supplemental Material). It is worth noting that high charge vortices, even those of a discrete nature, are difficult to realize in equilibrium quantum fluids since they tend to separate into smaller single-charge charge vortices which are more thermodynamically stable.

Scrutinizing the polarization of the OAM locked condensate solutions we are able to find rich patterns of polarization singularities which share strong similarities with electromagnetic textures known as optical skyrmions [87]. This offers perspectives on using interfering ballistic SOC polariton systems to generate lattices of polarization singularities and full Poincaré beams for singular spinoptronic technologies [46, 52]. A fu-ture perspective includes studying multistability and hysteresis [88] between our giant discrete polariton vortices in analogy to hysteresis found in e.g. magnetic skyrmions [89] and circulating currents in superconducting circuits [90].

The ability of the condensates to synchronize and phase-lock into a circulating current is bounded by the coherence length of the condensate array which decreases with increasing levels of disorder and condensate interactions with incoherent background excitons and charge carriers. We expect that disorder levels encountered in GaAs ($\sim$ $\mu$eV [91]) and single-crystal perovskite ($\sim$ meV [92]) can be controlled and minimized in our polariton vortex lasing arrays utilizing a feedback-stabilized multispot excitation scheme that has demonstrated coherence lengths in the order of $\sim$ 100 $\mu$m across various pumping geometries [82]. Another obstacle is that the TE-TM splitting should be comparable to the polariton linewidth in order for specific OAM lasing to become robust. This implies $\beta k_c^2 > \hbar\gamma/2$ where $k_c$ is the momentum of outflowing condensate polaritons. Using $2m\beta/\hbar^2 = 0.1$ and the same mass $m$ and decay rate $\gamma$ given in Table I we obtain a value of $k_c > 0.83$ $\mu$m$^{-1}$ which has been achieved in both GaAs ($k_c \approx 2$ $\mu$m$^{-1}$) [82] and $CsPbBr_3$ ($k_c \approx 4$ $\mu$m$^{-1}$) [82] cavities.

We believe that our results could be extended to novel room temperature polaritonic materials such as transition metal dichalcogenides with optically addressable spin-dependent energy shifts reported in $WSe_2/MoSe_2$ heterobilayer excitons [93] and $WS_2$ superlattice polaritons [94]. As mentioned, perovskite cavities [32, 62, 95, 96] are another promising platform for chiral optoelectronics [97] with recent demonstration of multiple coupled ballistic condensates in all-inorganic lead halide $CsPbBr_3$ [92] and flexible design for compact polariton lasing in circuit geometries [**?** ]. A possible future direction to further exploit our ballistic polariton laser is introducing chiral absorbing metasurfaces to the cavity design [6, 98, 99]. This would alleviate the need for strong TE-TM splitting but still benefit from the multiparameter optical control.

### Data availability

The datasets generated and analyzed during the current study are available from the corresponding author on reasonable request.

### Acknowledgments

The authors acknowledge the project No. 2022/45/P/ST3/00467 co-funded by the Polish National Science Centre and the European Union Framework Programme for Research and Innovation Horizon 2020 under the Marie Skłodowska-Curie grant agreement No. 945339. V.K.D. acknowledges the Icelandic Research Fund (Rannís), grant No. 239552-051. Z. W. acknowledges Erasmus+: Erasmus Mundus programme of the European Union. We thank Marcin Muszyński for assistance in drawing the cavity schematic in Fig. 1(d).

### Author contributions

Z.W., A.F., V.K.D., and H.S. performed the theoretical analysis and calculations. H.S. conceived the idea and supervised the project. Z.W. and H.S. wrote the manuscript with input from all other authors.

### Conflict of interest

The authors declare that they have no conflict of interest.

# Supplementary Material: Discrete chiral ballistic polariton laser

Zuzanna Werner,[1,*] Andrzej Frączak,[1] Valtýr Kári Daníelsson,[2]
Jacek Szczytko,[1] Barbara Piętka,[1] and Helgi Sigurðsson[1,2,†]

[1] *Institute of Experimental Physics, Faculty of Physics, University of Warsaw, ul. Pasteura 5, PL-02-093 Warsaw, Poland*
[2] *Science Institute, University of Iceland, Dunhagi 3, IS-107 Reykjavik, Iceland*

## S1. EXAMPLE CONDENSATE VORTEX SOLUTIONS ACROSS DIFFERENT POLYGONAL ORDERS

Figure S1 shows similar results to Fig. 8 in the main text but for different number of pump spots $N = 3, 5, 7, 9, 11$ in increasing polygonal order. As the number of spots increases, solutions with a higher vortex charge, given by $|\ell| \leq (N-1)/2$, become available as solutions of the system. Here, we use the same parameters as given in Table I in the main text and only vary slightly the pump ellipticity $\sin(2\Theta) > 0$ and polygon radius $R$. From top to bottom we display discrete vortex states in the dominantly pumped spin-up $\psi_+$ component from $\ell^+ = -1, 2, -3, -4, 5$. From left to right we show the normalized densities $|\psi_+|^2$ and $|\psi_-|^2$, and normalized $s_z$ Stokes (pseudospin) parameter. The Stokes phase $\phi_{x,y} = \arg(s_x + is_y)$ in the zoomed region outlined by a black dashed line is shown in the last column. The cyan arrows overlaid on the $|\psi_+|^2$ data show the phasor of each condensate. The These results underline the wide variety of vortex states available under our scheme.

The limit to the number of pump spots and the charge of the vortex is ultimately set by the coherence length of the polaritons which dictates their ability to synchronize into large scale patterns. Decoherence of polariton waves can originate from interactions with a background reservoir of excitons and charge carriers, phonon fluctuations, and sample disorder. However, coherence lengths in state of the art GaAs and perovskite cavities have reached 100 $\mu$m [1, 2] which is sufficient to capture a great variety of vortex states in our setup. We also expect that additional radial confinement (i.e., quantum rings) [3] will decrease the condensation threshold and reduce detrimental dephasing effects from photoexcited charge carriers and excitons.

## S2. CHOICE OF PARAMETERS IN 2DGPE SIMULATIONS

Most parameters in Table I used in our work are based on experimentally extracted and measured values for GaAs based cavities which served as motivation for this work [4]. The values of polariton lifetime $\gamma^{-1} \sim$ ps, effective mass $m$, and exciton Hopfield coefficient $|X|^2$ are typical for negatively detuned III-V cavities [5]. The exciton-exciton interaction strength $g_0 = 10$ $\mu$eV $\mu$m$^2$ has been verified with good precision in recent experiments [6]. The chosen SOC strength $2m\beta/\hbar^2 = 0.1$ is similar to what has been reported in experiments from Warsaw using liquid crystal microcavities [7].

The pump power $P_0$ is always chosen such that we are within $< 50\%$ above threshold in simulations which is similar to the approach in experiment [4] and reduces the chances of multiple stable solutions forming when nonlinear effects become too high. The pump spot width $w$, number of spots, and polygon radius $R$ is similar to what has been used in previous multispot polariton condensate experiments [1, 4, 8].

Parameters that are somewhat ambiguous are the stimulation (conversion) rate of excitons into polaritons $\Gamma_R$, the exciton decay (redistribution) rate $\Gamma_X$, the dark-bright reservoir population ratio $\eta$, and the exciton spin relaxation rate $\Gamma_s$.

(1) First, $\Gamma_R = 0.2$ meV$\mu$m$^2/\hbar$ can typically be found from matching simulated condensate patterns around threshold to those seen in experiment [9]. That is, the outflow $k$-vector of ballistic condensate polaritons determines their reciprocal space distribution and number of real space interference fringes between condensates. This $k$-vector is proportional to $\Gamma_R^{-1}$ since the latter dictates how large of a reservoir density (blueshift) $n_\pm \propto \gamma/\Gamma_R$ is needed to reach threshold. We have used a value of $\Gamma_R$ that reproduces typical interference patterns between condensates like seen in past experiments [1, 4]. (2) Second, $\Gamma_X = 0.1$ meV$/\hbar$ can in principle be much smaller than we have chosen here since excitons can have long nonradiative nanosecond lifetimes. However, it has been proposed that bottleneck excitons which feed the condensate have shorter lifetimes due their proximity to the light cone that enhances radiative losses [10]. CW-excitation experiments [9] have found good agreement taking $\Gamma_X \sim \gamma$ when focusing on steady state solutions instead of transient dynamics. (3) Third, the ratio between dark and bright exciton reservoirs $\eta$ is most difficult to predict given the optical inaccessibility of dark exciton states. However, a recent theoretical study suggests that $\eta\Gamma_R \sim 2$ (i.e., two dark excitons for every bright exciton) [11] gives good agreement with measurements on the sound velocity of a polariton fluid. Thus we have chosen this value. We note that $\eta$ is not necessary for the reported findings and we can set $\eta \to 0$ and obtain qualitatively the same patterns since substantial blueshift and optical Zeeman effect can originate just the reservoir of bright bottleneck excitons $n_\pm$. (4) Fourth, $\Gamma_s$ should generally be fast (several to tens of picoseconds) in GaAs based quantum wells [12] and lead halide perovskite crystals [13] which is what we have chosen here. Importantly, lower values of $\Gamma_s$ do not qualitatively affect our results since slower spin relaxation simply means more efficient pumping of a given polariton spin and triggering of OAM selection. Therefore, we can set $\Gamma_s \to 0$ and still obtain the same results presented in the study.

* z.werner@student.uw.edu.pl
† helgi.sigurdsson@fuw.edu.pl

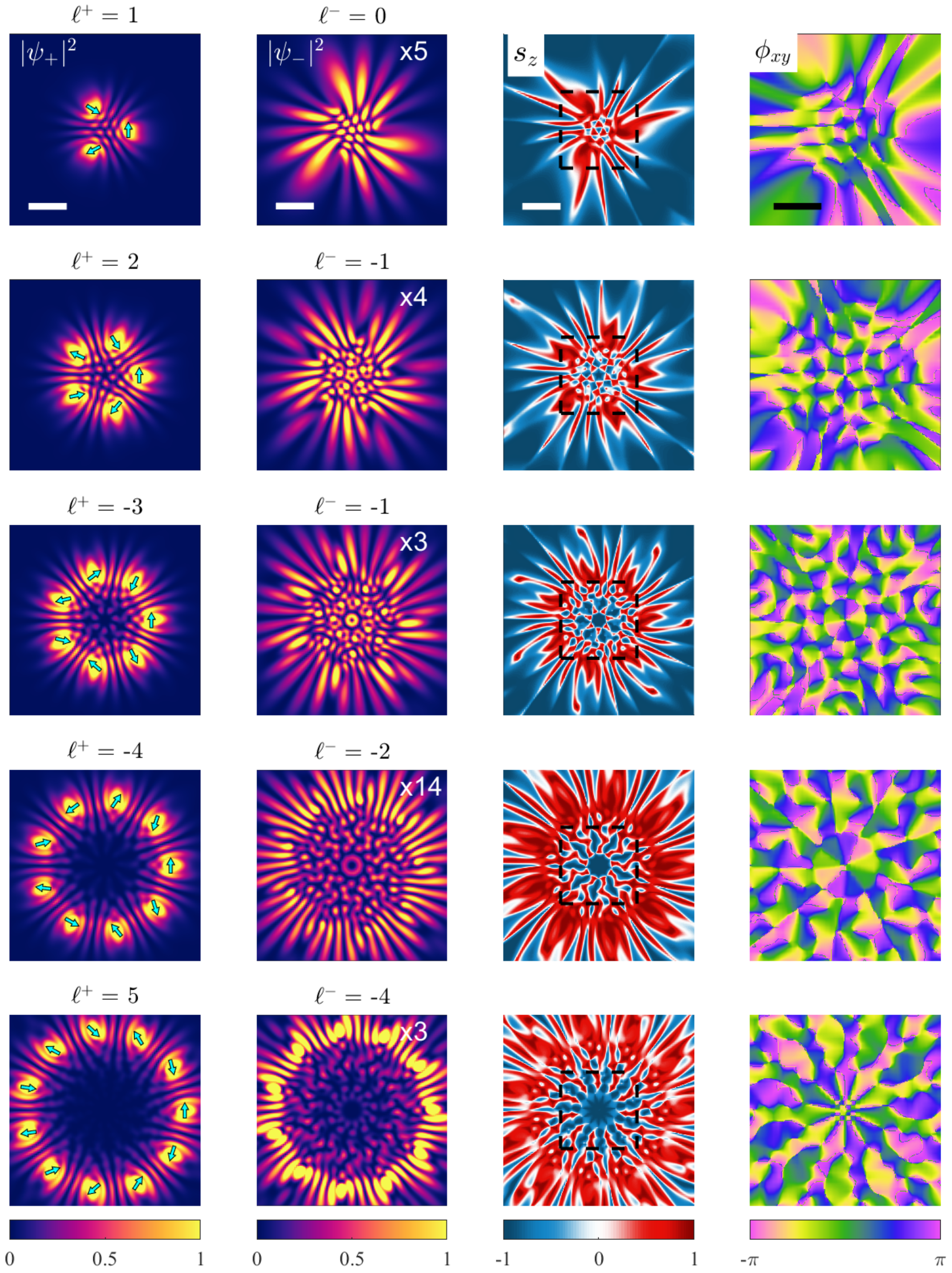

FIG. S1. Steady state giant vortex solutions obtained from spinor 2D GPE simulations. (Left to right panels) Condensate spin-up density $|\psi_+|^2$, spin-down density $|\psi_-|^2$, normalized degree of circular polarization $s_z$, and Stokes phase $\phi_{xy} = \arg(s_x + is_y)$. Like in Fig. 8 in the main text $|\psi_+|^2$ is dominantly pumped with $\Theta > 0$. (Top to bottom panels) Increasing number of pump spots (polygon vertices) $N = 3, 5, 7, 9, 11$ displaying different OAM $|\ell^+| = (N-1)/2$. The sign of $\ell^+$ depends on the polygon radius, pump power, and pump ellipticity like shown in the phase map of Fig. 7 in the main text. We multiply the data with a scalar (upper right corner) to increase visibility in the plots. White scale bars are 20 $\mu$m and black 10 $\mu$m.

In short, most parameters in our model accurately represent a typical polariton III-V Fabry-Perot cavity. The few parameters that are difficult to estimate are chosen within reasonable estimations of their values and we can take $\eta$, $\Gamma_s = 0$ without any loss of generality. The important point is that the results are not critically dependent on any specific choice of parameters as long as the pumping geometry is intact, SOC is present, and polariton interactions blueshift the spin levels different amounts (optical Zeeman effect). This is the general condition that leads to hybridization between different OAM carrying states in different angular Bloch bands. This condition can be upheld for a large range of reservoir parameters as long as steady state solutions exist and no specific energy terms dominates over other (e.g., the reservoir blueshift cannot be so large that it invalidates the parabolic approximation for polaritons).

---